\documentclass[%
 aip,
 jcp,
 amsmath,amssymb,
reprint,%
]{revtex4-1}

\usepackage{graphicx}
\usepackage{dcolumn}
\usepackage{bm}

\usepackage[utf8]{inputenc}
\usepackage[T1]{fontenc}
\usepackage{mathptmx}
\usepackage{etoolbox}

\makeatletter
\def\@email#1#2{%
 \endgroup
 \patchcmd{\titleblock@produce}
  {\frontmatter@RRAPformat}
  {\frontmatter@RRAPformat{\produce@RRAP{*#1\href{mailto:#2}{#2}}}\frontmatter@RRAPformat}
  {}{}
}%
\makeatother
\begin{document}


\title[]{Comparing dimensionality reduction methods for local structural identification in colloidal systems}

\author{A. Ulug\" ol}
 \affiliation{%
Soft Condensed Matter and Biophysics Group, Debye Institute for Nanomaterials Science, Utrecht University, Princetonplein 1, Utrecht, 3584 CC, Netherlands
}%
\email{a.ulugol@uu.nl.}

\author{J.I. B\"uckmann}
\affiliation{%
Soft Condensed Matter and Biophysics Group, Debye Institute for Nanomaterials Science, Utrecht University, Princetonplein 1, Utrecht, 3584 CC, Netherlands
}%

\author{R. Yang}
\affiliation{%
Soft Condensed Matter and Biophysics Group, Debye Institute for Nanomaterials Science, Utrecht University, Princetonplein 1, Utrecht, 3584 CC, Netherlands
}%

\author{L.D. Hoitink}
\affiliation{%
Soft Condensed Matter and Biophysics Group, Debye Institute for Nanomaterials Science, Utrecht University, Princetonplein 1, Utrecht, 3584 CC, Netherlands
}

\author{A. van Blaaderen}
\affiliation{%
Soft Condensed Matter and Biophysics Group, Debye Institute for Nanomaterials Science, Utrecht University, Princetonplein 1, Utrecht, 3584 CC, Netherlands
}%

\author{F. Smallenburg}
\affiliation{%
Universit\'e Paris-Saclay, CNRS, Laboratoire de Physique des Solides, 91405 Orsay, France 
}%

\author{L. Filion}
\affiliation{%
Soft Condensed Matter and Biophysics Group, Debye Institute for Nanomaterials Science, Utrecht University, Princetonplein 1, Utrecht, 3584 CC, Netherlands
}%

\date{\today}

\begin{abstract}
Quantifying local structures in self-assembled systems is a central challenge in soft matter and materials science. When no \emph{a priori} knowledge of the relevant structures is available, traditional order parameters often fall short. Unsupervised machine learning provides a convenient route to autonomously uncover structural motifs directly from particle configurations. In this work, we systematically compare three popular dimensionality reduction techniques; Principal Component Analysis (PCA), Autoencoders (AE), and Uniform Manifold Approximation and Projection (UMAP), for classifying local environments in self-assembled systems. We first apply these methods to fluid and crystal configurations of hard and charged spheres. Thereafter, we apply it to an icosahedral arrangement of spheres that self-assembled in spherical confinement, both from simulations as well as from experiments. We demonstrate that UMAP consistently outperforms the other methods in capturing complex structural features, offering a robust tool for structural classification without supervision.
\end{abstract}

\maketitle

\section{Introduction}
Understanding the structural organization of self-assembled materials is a fundamental problem in soft matter, condensed matter, and materials science. A key challenge lies in identifying and classifying the local environments that particles occupy, particularly in cases where the global symmetry is imperfect, multiple polymorphs coexist, or the relevant structural motifs are not known in advance. Traditional approaches rely on manually crafted order parameters tailored to known crystal structures\cite{auer2001prediction, gasser2001real, lechner2008accurate, hermes2011nucleation, sanz2008out, de2023search, de2024phase,van1995real}, but these can fail when unexpected or complex arrangements arise.

In the last decade, alternative approaches based on machine learning have been introduced, both from a supervised and unsupervised approach.  In the case where the potential local environments are known beforehand, supervised machine learning can be used to train classifiers to recognize these different options (see e.g. Refs. \onlinecite{geiger2013neural,boattini2018neural,coli2021artificial, spellings2018machine, coli2022inverse}). However, in the case where there is no \emph{a priori} set of structures to look for, unsupervised machine learning (UML) can be a powerful alternative.

A key strength of UML is its ability to extract structural patterns directly from the data without requiring labeled input. Recent studies have demonstrated the potential of UML techniques to detect subtle structural heterogeneities in systems ranging from glasses\cite{boattini2020autonomously,alkemade2023improving,jung2023predicting,boattini2021averaging,coslovich2022dimensionality,qiu2025unsupervised,oyama2023deep,shiba2023botan,alkemade2022comparing} to colloidal crystals\cite{leoni2021nonclassical,lotito2020pattern,clegg2021characterising,adorf2019analysis,telari2023charting,chapman2023quantifying,o2021deep,becker2022unsupervised,reinhart2021unsupervised,kyvala2025unsupervised,adorf2019analysis,bedolla2025relationship}, often uncovering hidden order that escapes conventional analysis.  In many cases these methods combine three components: local structural descriptors, dimensionality reduction to reveal key variations in the data, and clustering to assign particles to distinct environments.

In this work, we systematically investigate the performance of three distinct dimensionality reduction techniques that have been used in the soft matter literature for the unsupervised classification of local structures in complex crystalline systems: principal component analysis (PCA)\cite{pearson1901liii}, autoencoders (AE)\cite{lecun1987connexionist}, and uniform manifold approximation and projection (UMAP)\cite{mcinnes2018umap}. Our goal is to evaluate how effectively each method captures structural diversity and separates distinct environments in a fully unsupervised pipeline.

The article is organized as follows. In Section~\ref{sec:methods}, we describe the overall methodology, detailing the structural descriptors, dimensionality reduction algorithms, and clustering techniques employed. Section~\ref{sec:results} presents a comparative evaluation of PCA, AE, and UMAP on two test sets of simulated self-assembled systems: (i) bulk crystalline structures as benchmark against labeled data and (ii) spherically confined hard spheres which form icosahedrally symmetric suprastructures as a complex classification problem. Note that to show the transferability of the method, we also show the analysis on one experimental dataset. 

Various descriptor families have been proposed for characterizing local structure in condensed matter systems, ranging from traditional bond-orientational order parameters\cite{steinhardt1983bond,lechner2008accurate} to more expressive representations such as Smooth Overlap of Atomic Positions (SOAP)\cite{bartok2013representing,de2016comparing,rapetti2023machine}, Local Environments and Neighbors Shuffling (LENS)\cite{crippa2023detecting,caruso2025classification} and Point Group Order Parameters (PGOPs)\cite{fijan2025quantifying}. Recent studies have shown that the choice of descriptor, as well as its interaction with a chosen embedding or clustering method, can influence the resolution and sensitivity of unsupervised classifications\cite{martino2025data,lionello2025relevant}. In this work, we do not compare descriptor families or clustering algorithms. Instead, we focus on the more limited but well-defined question of how different dimensionality-reduction approaches perform when used as interchangeable components within the same unsupervised workflow, where both the structural descriptor set and the classification step are kept consistent. Our results should therefore be interpreted as a comparison of embedding choices within a fixed workflow rather than a general benchmark of unsupervised learning methods.

\section{Methods}\label{sec:methods}
This section outlines the full unsupervised pipeline used to classify local particle environments. Our approach proceeds in three main stages, similar to the approach introduced in Ref.~\onlinecite{boattini2019unsupervised}. We begin by encoding the local structure around each particle into a set of numerical descriptors. These high-dimensional feature vectors are then projected into a low-dimensional space using dimensionality reduction techniques. Finally, clustering is performed in this reduced space to assign particles to distinct structural classes.

\subsection{Local Structure Descriptors}\label{subsec:descriptors}
The first step is to quantify the local environment of each particle in a way that captures relevant structural information. To this end, we use two families of descriptors: rotation-invariant bond orientational order parameters\cite{lechner2008accurate,steinhardt1983bond,auer2005numerical}, and a scalar measure of local density asymmetry that we introduce here.


To characterize local angular order, we first determine a set of nearest neighbors for each particle $i$ using the solid-angle–based nearest-neighbor (SANN) criterion \cite{van2012parameter}. Unlike radial cutoff methods, which require choosing a fixed analysis radius, SANN assigns each particle an adaptive, particle-specific radius $r_\mathrm{SANN}(i)$ such that the solid angles of its neighbors exactly fill that of a sphere. All particles within this radius are considered neighbors\cite{van2012parameter}. Recent work has shown that bond-orientational order parameters can depend sensitively on the choice of cutoff distance\cite{doria2025data,martino2025data}. In this study we therefore use SANN as a parameter-free, geometry-based neighbor definition, avoiding the need to select an explicit radial cutoff for the computation of local order parameters.

Next, we compute complex bond orientational order parameters defined for particle $i$ at position $\mathbf{r}_i$ by\cite{steinhardt1983bond}:
\begin{equation}
    q_{\ell}^m(i) = \frac{1}{\left| \mathcal{N}(i) \right|}\sum_{j\in\mathcal{N}(i)} Y_\ell^m
    (\hat{\mathbf{r}}_{ij}),
\end{equation}
where $Y_\ell^m$ are the spherical harmonics, $\hat{\mathbf{r}}_{ij}$ is the unit vector from particle $i$ to its neighbor $j$, and $\mathcal{N}(i)$ denotes the set of nearest neighbors of particle $i$ (with $\left| \mathcal{N}(i) \right|$ its size). Following Ref. \onlinecite{lechner2008accurate}, we then define a locally averaged bond order parameter via:
\begin{equation}
    \overline{q}_{\ell}^m(i) = \frac{1}{1+\left| \mathcal{N}(i) \right|}\sum_{k\in \{i\}\cup\mathcal{N}(i)}q_{\ell}^m(k).
\end{equation}
From these complex coefficients, we compute the scalar rotation-invariant order parameter\cite{lechner2008accurate}:
\begin{equation}
    \overline{q}_\ell(i) = \sqrt{\frac{4\pi}{2\ell + 1} \sum_{m=-\ell}^\ell \left|\overline{q}_{\ell}^m(i)\right|^2}.
\end{equation}
In this paper, we consider bond order parameters with $\ell$ ranging from $1$ to $12$.

To complement these angular descriptors with a measure of positional symmetry, we define the center of mass of the nearest neighbors as:
\begin{equation}
    \mathbf{r}_\mathrm{NNC}(i) = \frac{1}{\left| \mathcal{N}(i) \right|}\sum_{j\in\mathcal{N}(i)} \mathbf{r}_{ij}
\end{equation}
The normalized distance between the particle and this center is given by:
\begin{equation}
    \delta r(i) = \frac{|\mathbf{r}_i - \mathbf{r}_\mathrm{NNC}(i)|}{r_\mathrm{SANN}(i)}. 
\end{equation}
This scalar parameter captures deviations from local centrosymmetry and can help distinguish highly ordered environments from asymmetric or interfacial ones.

Together, these descriptors form a $13$ dimensional feature vector that encodes the local structure of each particle.

\subsection{Dimensionality reduction}

The local structure descriptors introduced above form high-dimensional feature vectors that may contain redundant or irrelevant information. To extract the dominant structural variations and facilitate clustering, we project these descriptors onto a lower-dimensional manifold. This step is used to highlight the most significant modes of variation in the data and to reduce sensitivity to small perturbations in the high-dimensional feature space\cite{vincent2008extracting,jolliffe2016principal,mcinnes2018umap}.  We compare three widely used techniques for this task: principal component analysis (PCA), autoencoders (AE), and uniform manifold approximation and projection (UMAP).

\subsubsection{Principal component analysis (PCA)}

PCA is a linear dimensionality reduction technique that identifies directions in feature space along which the variance of the data is maximized \cite{pearson1901liii}. It achieves this by computing the eigenvectors of the covariance matrix of the input data and projecting each data point onto the leading eigenvectors (principal components). The result is a lower-dimensional representation that captures as much of the total variance as possible under a linear transformation.

In physical terms, the first principal component corresponds to the direction of maximum structural variability across particles, the second component captures the next most significant variation orthogonal to the first, and so on. PCA is fast, interpretable, and the only hyperparameter is the dimensionality of the projected data. However, its reliance on linear combinations limits its ability to capture nonlinear structural correlations that could be present in complex systems.

\subsubsection{Autoencoder (AE)}

Autoencoders are a class of artificial neural networks designed to learn efficient representations of input data\cite{kramer1991nonlinear,baldi1989neural}. An autoencoder consists of two parts: an encoder that maps high-dimensional inputs to a low-dimensional latent space (the bottleneck), and a decoder that attempts to reconstruct the original inputs from this latent representation. By training the network to minimize the reconstruction error between input and output, the encoder learns a nonlinear mapping that compresses the data while retaining its most salient features. Unlike PCA, the autoencoder can learn complex, nonlinear relationships between descriptors. However, in addition to the dimensionality of the reduced space (which is set by the size of the bottleneck layer), AE has a variety of hyperparameters to tune such as hidden layer count and width, activation functions, learning rate, and weight regularizers. 

While extensive tuning of these parameters may further improve performance, it comes with two important caveats: (i) the optimization process becomes increasingly time-consuming, and (ii) excessive tuning increases the risk of overfitting, particularly in the absence of labeled data. To mitigate this risk, a portion of the data should be withheld for validation. However, in the context of unsupervised learning, where true structural labels are unavailable, defining a representative and balanced train–test split becomes nontrivial. In particular, care must be taken to ensure that all relevant local structures are sufficiently represented in both subsets to enable meaningful performance assessment.

In this work, we use a simple feedforward architecture with two hidden layers in both the encoder and the decoder as presented in Ref.~\onlinecite{boattini2019unsupervised}. Once trained, we discard the decoder and retain the encoder to project each particle’s structural descriptor into the latent space. 

To assess whether this minimal architecture underestimates the potential of autoencoders, we systematically explore deeper network architectures in the Supplementary Material (SM) by increasing the number of hidden layers while keeping all other training parameters fixed. We find that increasing network depth leads to a considerable reduction in reconstruction loss (up to $\sim25\%$), indicating improved representational capacity. By contrast, improvements in neighborhood preservation between the original space and the latent space are comparatively small, and the resulting cluster structure shows only modest refinement. These results indicate that while deeper autoencoders better reproduce the input descriptors, the qualitative conclusions drawn from the autoencoder analysis in the main text are robust and not an artifact of an under-capacitated model.

\subsubsection{Uniform manifold approximation and projection (UMAP)}
\begin{figure}
    \centering\includegraphics[width=\columnwidth]{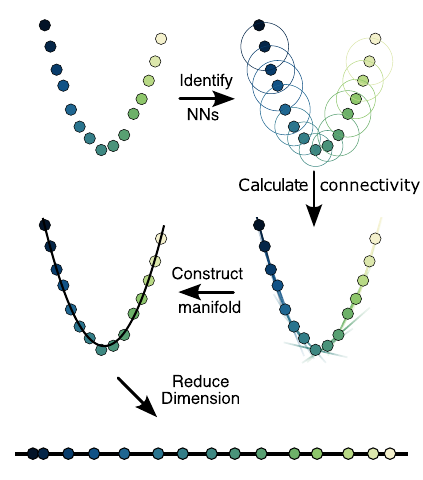}
    \caption{Schematic representation of UMAP's working principle.}
    \label{fig:umap-proj}
\end{figure}

UMAP is a nonlinear dimensionality reduction algorithm introduced in 2018 \cite{mcinnes2018umap} that constructs a low-dimensional representation of high-dimensional data by modeling its intrinsic geometry. 

At its core, UMAP assumes that the data lies on a low-dimensional Riemannian manifold embedded in a higher-dimensional space. The goal is to recover a faithful projection of this manifold in a lower-dimensional space (typically 2D or 3D) while preserving both local and global geometric features.

The algorithm, illustrated in Fig.~\ref{fig:umap-proj}, proceeds in two main steps:

\paragraph{Local neighborhood construction in high dimensions.}

For each particle, UMAP identifies  nearest neighbors within the high-dimensional space (here given by our chosen set of descriptors). It then encodes the local connectivity into a weighted graph where the weight $w_{ij}$ between points $i$ and $j$ reflects how close they are in the original space. The specific choice of weights used in this graph derives from a theoretical framework based on Riemannian geometry and algebraic
topology.
One key to UMAP is the way it generates this graph, which is based on a fuzzy topological representation.  For more information, see Ref. \onlinecite{mcinnes2018umap}.

\paragraph{Low-dimensional projection via optimization.}

Once local connectivity is modeled in the high-dimensional space, UMAP constructs an analogous set of points in the lower-dimensional space, and optimizes it such that it gives rise to the graph weights that are close to those for the higher-dimensional dataset.
Specifically, the projection $\mathbf{y}_i$ is obtained by minimizing a cross-entropy–based loss function between the two fuzzy sets. The associated loss function is:
\begin{equation}
\mathcal{L} = \sum_{i \ne j} \left[ w_{ij} \log \left( \frac{w_{ij}}{w_{ij}^\prime} \right) + (1 - w_{ij}) \log \left( \frac{1 - w_{ij}}{1 - w_{ij}^\prime} \right) \right],
\end{equation}
where $w_{ij}$ denotes the edge weight between points $i$ and $j$ in the high-dimensional space, and $w_{ij}^{\prime}$ is the corresponding weight in the low-dimensional projection. The weights are constructed using smooth exponential or inverse distance kernels, with parameters chosen to balance local vs. global structure.

This loss function measures the cross-entropy between the two neighborhood graphs, and optimization proceeds via stochastic gradient descent. This process seeks a projection that maximizes the agreement between neighborhood structures, or equivalently, preserves as much of the relational information from the original space as possible.

This perspective aligns UMAP with the broader class of manifold learning methods, where the goal is to retain the intrinsic geometry of the data rather than its precise coordinates\cite{hinton2002stochastic,maaten2008visualizing,tenenbaum2000global,kohonen1982self}. The probabilistic graph formulation also makes UMAP robust to noise and capable of capturing subtle topological features in structural datasets.

\paragraph{Choice of hyperparameters.}

UMAP exposes two primary hyperparameters that control the balance between local and global structure in the embedding: the number of nearest neighbors (\#NN) used to construct the high-dimensional graph, and the minimum distance between embedded points ($d_\mathrm{min}$) in the low-dimensional space. The former determines the size of the local neighborhood whose geometry UMAP prioritizes, while the latter influences how tightly clusters may form in the projection. A detailed comparison of the effects of the hyperparameters can be found in Ref. \onlinecite{mcinnes2018umap}. In practice, we find that our structural classifications are robust to variation in these parameters within commonly used ranges. Changing \#NN between $10$ and $100$ or $d_\mathrm{min}$ between $10^{-4}$ and $0.1$ does not alter the qualitative structure of the embedding nor the resulting clustering. Throughout this work, we therefore use a single consistent set of UMAP hyperparameters for all datasets.

Once the UMAP model is trained, the model stores both the high- and low-dimensional graphs. To transform a point outside of its training dataset, UMAP adds the new data into its stored high-dimensional graph, constructs its local neighborhood and weights, then reoptimizes its loss function by tuning the transformed coordinates of the new data.

\subsection{Clustering}

After projecting the high-dimensional structural descriptors into a low-dimensional space, we identify distinct structural classes by applying a clustering algorithm. We use a Gaussian Mixture Model (GMM)\cite{mclachlan2000finite, fraley2002model} to model the distribution of particles in this reduced space as a weighted sum of multivariate Gaussian components. The GMM assigns each particle a probability of belonging to each component, enabling both hard and soft classification.

A key challenge in unsupervised classification is determining the appropriate number of clusters, $N_c$, for Gaussian mixture models (GMMs). To address this, we fit models with varying values of $N_c$ and evaluate the model quality using the Bayesian Information Criterion (BIC)\cite{schwarz1978estimating} and the Akaike Information Criterion (AIC)\cite{akaike2003new}:
\begin{equation}
\begin{aligned}
\mathrm{AIC} &= 2k - 2 \log \mathcal{L},\\
\mathrm{BIC} &= k\log N  - 2 \log \mathcal{L},\
\end{aligned}
\end{equation}
where $\mathcal{L}$ is the maximum likelihood of the model, $k$ is the number of free parameters, and $N$ is the number of data points. Both criteria penalize model complexity to avoid overfitting, but BIC applies a stronger penalty for large datasets and is more conservative in selecting complex models that have a large number of free parameters.

Rather than strictly choosing the cluster number that minimizes BIC or AIC, we typically identify the point of diminishing return, where adding more components no longer significantly improves model likelihood. This reflects the practical reality that the true structure of the data is often not well described by a mixture of Gaussians: local environments may be non-Gaussian, anisotropic, or smoothly varying. In such cases, strict minimization of BIC can lead to oversegmentation, where a physically meaningful cluster is fragmented across multiple GMM components.

Additionally, we implement a post-processing step that iteratively merges GMM components based on their entropy.
To this end, we apply the entropy-based merging scheme introduced by Baudry et al.~\cite{baudry2010combining}, which constructs a hierarchy of candidate clusterings by successively merging the pair of components whose union yields the largest decrease in entropy:
\begin{equation}
S = -\sum_{i=1}^{N} \sum_{j=1}^{K} p_{ij} \log p_{ij},
\end{equation}
where $p_{ij}$ is the posterior probability that particle $i$ belongs to component $j$. This procedure yields a sequence of models with decreasing cluster number $N_c$, from which we identify the optimal clustering via visual inspection guided by the L-method, i.e., by detecting the ``elbow'' point in the entropy-vs-$N_c$ curve\cite{baudry2010combining}.

\subsection{Experimental}
Experimental icosahedral supraparticles were prepared in order to demonstrate the unsupervised ML algorithm on an experimental data set, similar to the icosahedral clusters. This was done by synthesizing fluorescent core-shell silica particles, which, once assembled, were analyzed using 3D stimulated emission depletion (STED) confocal microscopy. In the following, we outline how these supraparticles, and their building blocks, were made and characterized.

\subsubsection{Supraparticle formation}
The particles were made in three steps, where small silica seeds are created in the first step, to provide a reasonably monodisperse starting point. In the second step, a fluorescent shell was grown around the particles, through which the particles could be detected with confocal/STED microscopy. Finally, a larger non-fluorescent silica shell was grown around the particles to ensure that the fluorescent domains were well-separated. The initial silica seeds were prepared in an amino acid stabilized solution, based on the procedure presented by Shahabi et al.~\cite{ShahabiAACS}. This procedure was followed to obtain 63 nm $\pm$ 4.5 \% silica seeds. The particles were then grown to 123 nm $\pm$ 2.6 \% using a seeded growth St\"ober method~\cite{vanBlaaderen1992three}. In this method, fluorescein isothiocyanate (FITC) was incorporated through the silance coupling agents (3-aminopropyl)triethoxysilane (APTES)~\cite{vanBlaaderen1992,Verhaegh}. Finally, in two additional steps the particles were grown to 326 nm $\pm$ 0.8 \% and 478 nm $\pm$ 0.8 \% using a seeded growth St\"ober method. The self-assembly of these particles was done following the simplified double-emulsion method reported by Wang \textit{et al.}~\cite{WangDE}.

\subsubsection{Supraparticle Analysis}
After several washing steps to remove surfactants, all supraparticles were dried on a \#1.5H high precision cover slip (Menzel Gl\"aser), glued to a standard microscopy slide (Menzel Gl\"aser) in which a 8 mm hole was drilled. The supraparticles were index-matched within 0.002 using a mixture of 82.5 wt\% glycerol/water.
All confocal data was acquired with a Leica TCS SP8 3x inverted confocal laser scanning microscopy using a 93x/1.3 HC PL APO CS2 STED objective (Leica \#506417) in the Leica Application Suite (LAS) X software (v3.5.6). A white-light laser (SuperK) was set to 488 nm with an intensity of 3\% to excite the particles. A 592 nm continuous-wave depletion laser (Leica STED) was used with a intensity of 30\%, and 100\% 3D STED was used to optimize the resolution in the axial direction. The resulting fluorescence was detected on a hybrid detector between 498 and 582 nm with a gating time of 0.3-12 ns. This was done in 16-bit depth for 4 line accumulations, a scan speed of 200 Hz and a typical $x\times y\times z$ pixel size of 30$\times$30$\times$80 nm$^{3}$. 
After acquisition, the data was deconvolved using Huygens Professional (SVI, V23.04), the individual particles were located using Trackpy (V0.6.3 \cite{Trackpyversion,Crockertracking}), and the clustering procedure was performed. 

\section{Results}\label{sec:results}

To benchmark the performance of the dimensionality reduction techniques, we first test their ability to distinguish between known crystalline structures. Subsequently, we apply the same techniques to reveal complex self-assembled order in spherically confined hard spheres. In each case, we assess the quality of the clustering and analyze the autonomously discovered structural motifs.

\subsection{Bulk Crystalline Structures}

\begin{figure}
    \centering
    \includegraphics[width=\linewidth]{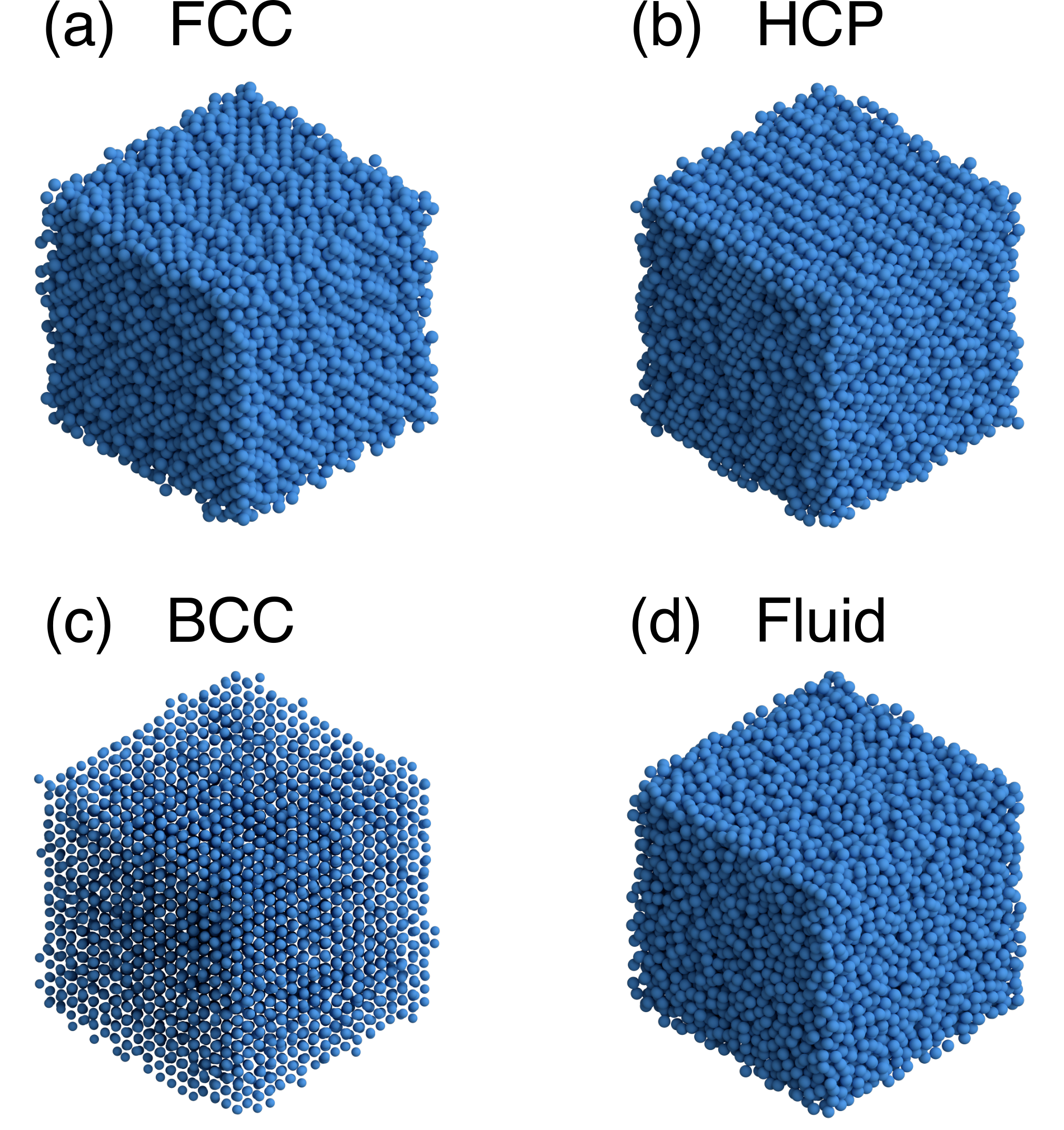}
    \caption{The dataset used in the bulk structure analysis.}
    \label{fig:bulk}
\end{figure}

As a first test, we generate a reference dataset by performing NVT Monte Carlo simulations of approximately $11000$ particles in different phases: face-centered cubic (FCC), hexagonal close-packed (HCP), body-centered cubic (BCC), and fluid phases. We use hard spheres to generate the FCC, HCP, and fluid phases, while the BCC phase was generated in a point Yukawa system. The snapshots of these systems are shown in Fig. \ref{fig:bulk}.

\begin{figure}
    \centering
    \includegraphics[width=\linewidth]{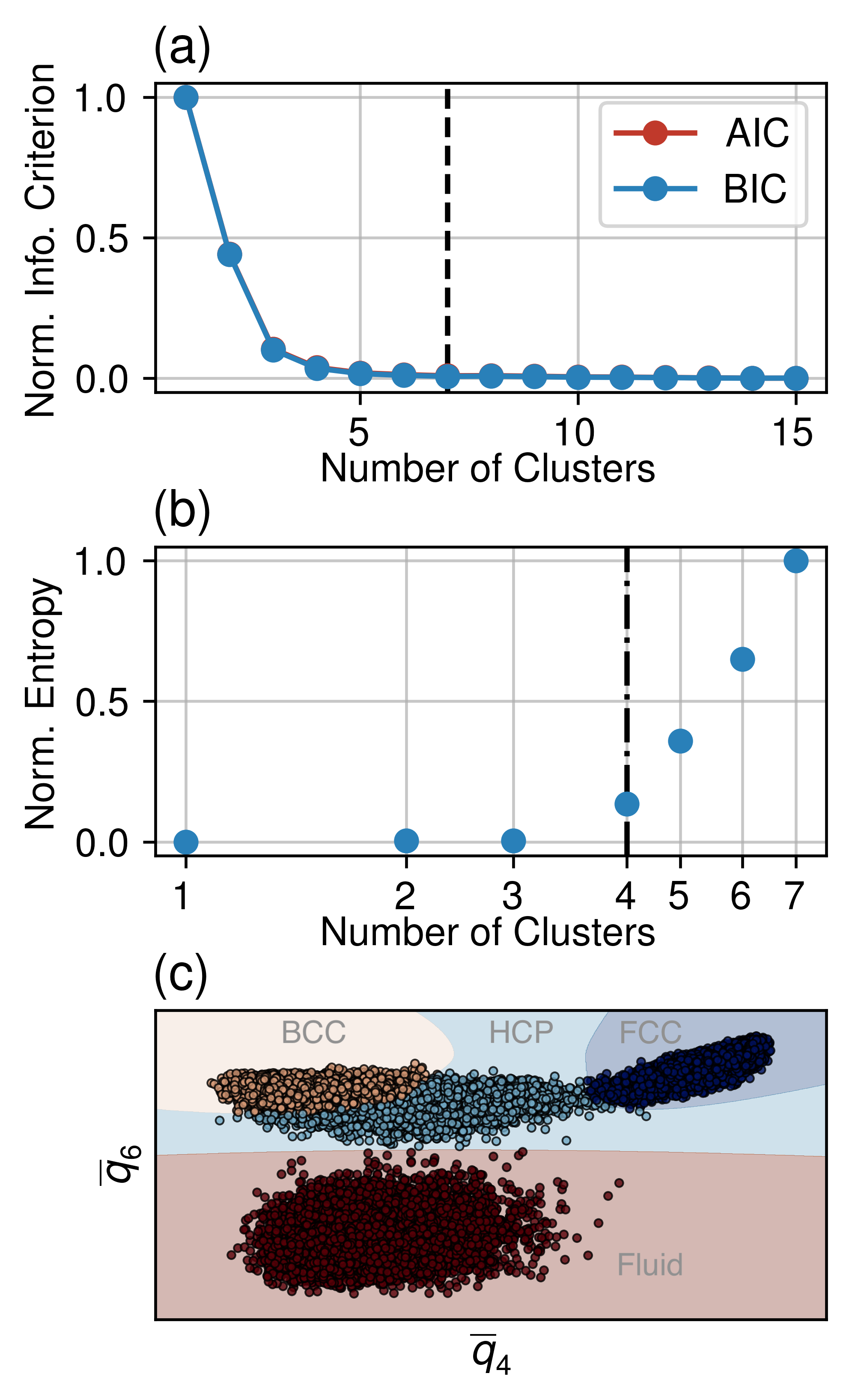}
   
    \caption{Clustering of bulk structures in the $\overline{q}_4$ vs. $\overline{q}_6$ plane. 
    (a) AIC and BIC scores for GMM clustering. (b) Entropy as a function of the number of clusters after successive merging.
    (c) Projection and classification after entropy-based merging into $4$ clusters. Note that the labels  (FCC, HCP, BCC and fluid) in this plot are based our a priori knowledge of the phases associated with each point, and not a result of the clustering procedure.}
    \label{fig:q4q6aicbic}
\end{figure}

As a baseline, we first employ the conventional approach based on bond orientational order parameters, namely the nearest-neighbor averaged $\overline{q}_4$ and $\overline{q}_6$\cite{lechner2008accurate, auer2005numerical}. We fit GMMs with up to 15 clusters to the resulting two-dimensional dataset. As shown in Fig.~\ref{fig:q4q6aicbic}(a), the Akaike Information Criterion (AIC) and Bayesian Information Criterion (BIC) nearly overlap, effectively collapsing onto a single curve. This strong overlap suggests that the likelihood term dominates both criteria, making the difference in their penalization schemes negligible. Notably, this behavior is not unique to this dataset, it recurs throughout our analyses. To avoid redundancy, we do not comment on it again for each case. Both AIC and BIC  decay rapidly and level out somewhere between 5 and 7 components. Conservatively, we choose 7 components for our initial model. Note that if the number of clusters chosen is too high, the entropy-based cluster merging procedure described in Section~\ref{sec:methods} will reduce the number of clusters. Hence, we typically err on the side of large numbers of clusters at this stage of the analysis. As shown in Fig.~\ref{fig:q4q6aicbic}(b), the elbow point in the entropy curve associate with the cluster merging occurs at 4 clusters, reflecting the number of underlying structures. We would like to point out that whenever we plot the information entropy against number of clusters, we scale the spacing of successive numbers of clusters proportional to the number of data points involved in the merging as suggested by Ref.~\onlinecite{baudry2010combining}.   

The final clustering is visualized in Fig.~\ref{fig:q4q6aicbic}(c). Several important observations emerge. First, the fluid phase (red) is well-separated from the crystalline phases, consistent with its lower bond orientational order. The FCC phase (dark blue) also forms a distinct, compact cluster. In contrast, the HCP (light blue) and BCC (orange) clusters show considerable overlap, reflecting the difficulty of distinguishing these structures based solely on these two bond orientational parameters.

These results establish a baseline against which the UML methods can be compared. 

\subsubsection{Principal component analysis}

Next, we assess the performance of PCA as a dimensionality reduction technique for the classification of bulk crystalline structures.

\begin{figure}
\centering
\includegraphics[width=\linewidth]{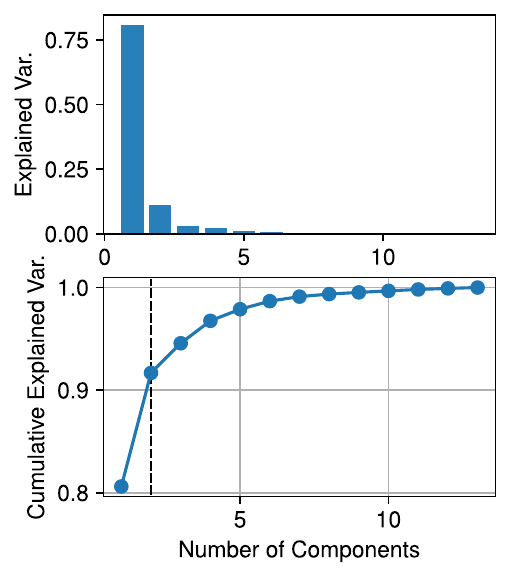}
\caption{Dimensionality analysis of bulk structures. (a) Variance explained by individual principal components and (b) cumulative explained variance.}
\label{fig:bulk_pca_dims}
\end{figure}

We first analyze the variance explained by each principal component. As shown in Fig.\ref{fig:bulk_pca_dims}(a), the first two principal components account for a substantial fraction of the total variance. The cumulative explained variance, plotted in Fig.\ref{fig:bulk_pca_dims}(b), confirms that more than $90\%$ of the structural variability is captured by the first two components. Therefore, projecting the data onto two dimensions provides a faithful representation while significantly reducing complexity.

\begin{figure}
\centering
\includegraphics[width=\linewidth]{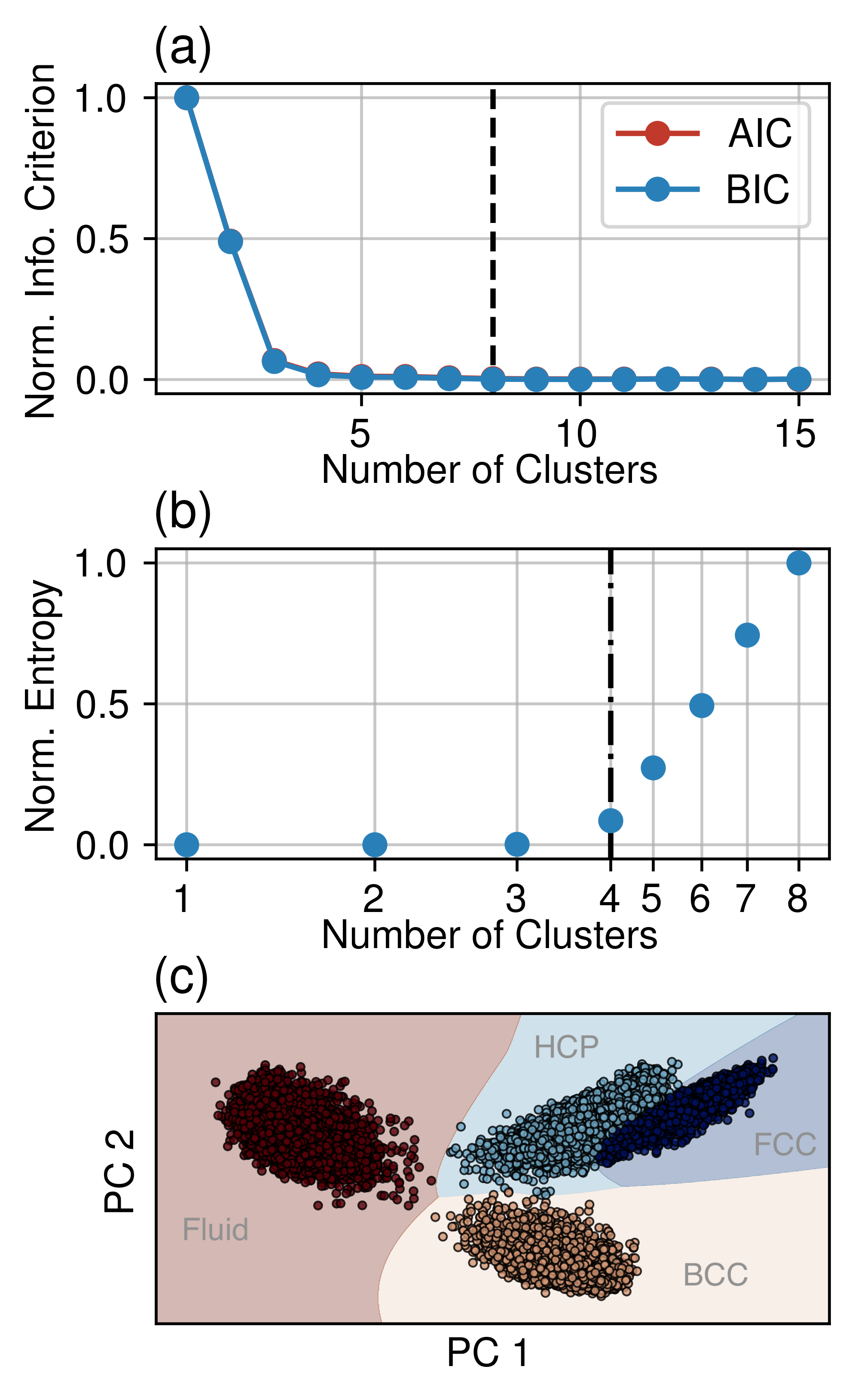}
\caption{Clustering of bulk structures in the PCA projected plane. 
    (a) AIC and BIC scores for GMM clustering. (b) Entropy as a function of the number of clusters after successive merging.
    (c) Projection and classification after entropy-based merging into $4$ clusters.}
\label{fig:bulk_pca_criterion}
\end{figure}

After projecting the data onto the first two principal components, we apply GMM-based clustering. Again, we fit models with up to 15 Gaussians and evaluate their quality using the AIC and BIC criteria (see Fig.~\ref{fig:bulk_pca_criterion}(a)). Conservatively, we choose a model with 8 Gaussians as a starting point for the cluster merging procedure. The cluster merging procedure again shows an elbow at four clusters, as shown in 
Fig.~\ref{fig:bulk_pca_criterion}(b), and hence we keep 4 clusters in the end.

The final classification is visualized in Fig.~\ref{fig:bulk_pca_criterion}(c). The fluid phase and the BCC structure are well-separated in the principal component space. However, similar to the $\overline{q}_4$ vs. $\overline{q}_6$ baseline analysis, PCA suffers from overlapping clusters. However, this time we observe significant overlap between the FCC and HCP phases.

\subsubsection{Autoencoders}

\begin{figure}
\centering
\includegraphics[width=\linewidth]{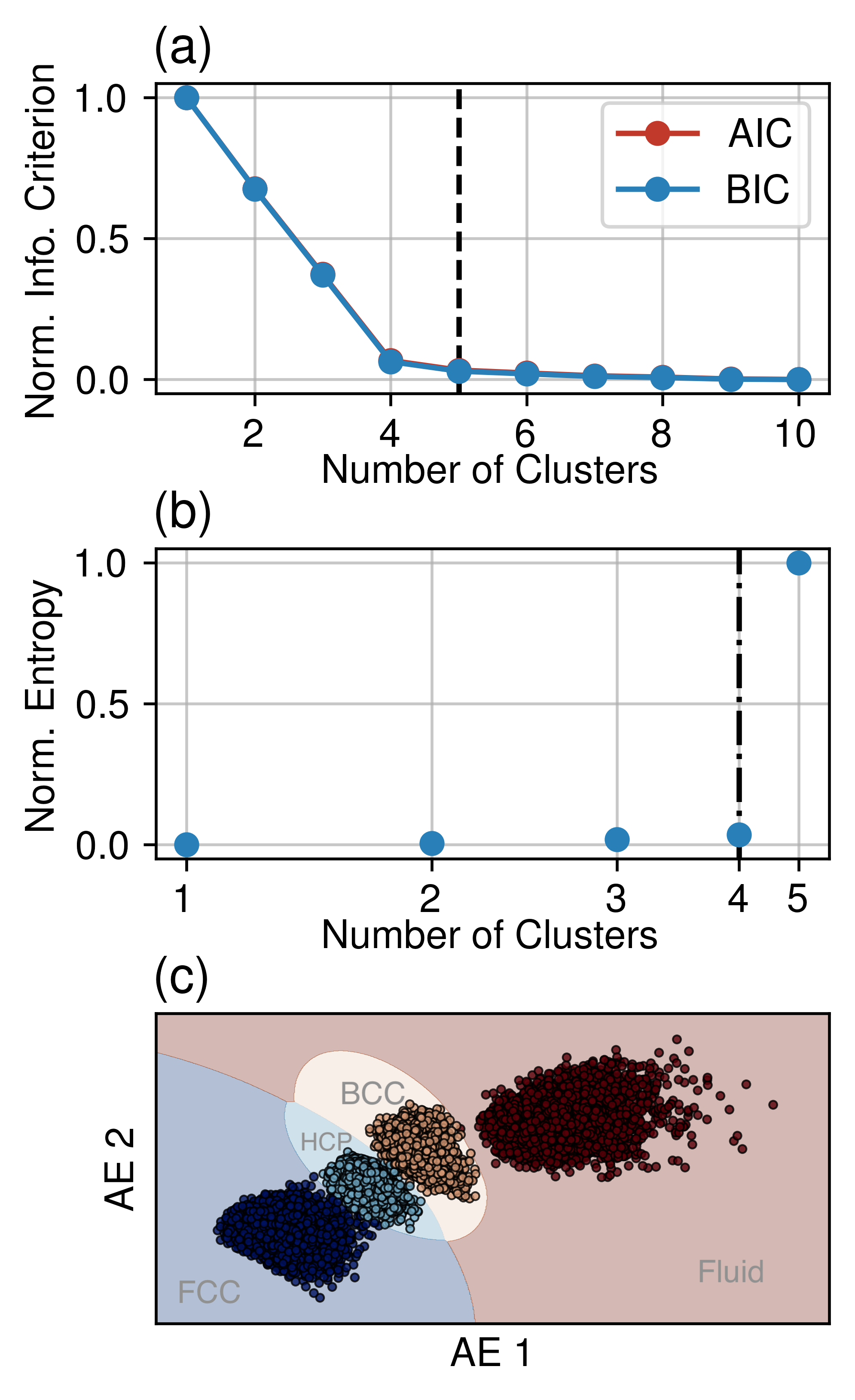}
\caption{Clustering of bulk structures in the autoencoder projected plane. 
    (a) AIC and BIC scores for GMM clustering. (b) Entropy as a function of the number of clusters after successive merging.
    (c) Projection and classification after entropy-based merging into $4$ clusters.}
\label{fig:bulk_ae_criterion}
\end{figure}

We now turn to autoencoders (AE) as a nonlinear dimensionality reduction technique for structural classification. Motivated by the results of the PCA analysis, we design the autoencoder to project the structural descriptors into a two-dimensional latent space.
After training the network to minimize reconstruction loss, we project the dataset into the learned two-dimensional bottleneck representation. We then apply GMM-based clustering to this latent space, varying the number of components up to $10$. As shown in Fig.~\ref{fig:bulk_ae_criterion}(a), both the AIC and BIC scores display a marked point of diminishing returns around 5 components, suggesting that finer substructure beyond this point does not substantially improve the model fit.
To refine the clustering, we again apply entropy-based component merging. As shown in Fig.~\ref{fig:bulk_ae_criterion}(b), the elbow point in the entropy curve occurs at 4 clusters, in agreement with the number of phases we simulated.

The final classification is shown in Fig.~\ref{fig:bulk_ae_criterion}(c). Unlike the results obtained using PCA or $\overline{q}_4$ vs. $\overline{q}_6$, the autoencoder successfully separates all four structural classes with minimal overlap. In particular, the fluid, FCC, HCP, and BCC clusters are all compact and well-separated, while remaining relatively close to one another in the latent space. 

\subsubsection{Uniform Manifold Approximation and Projection}

\begin{figure}
\centering
\includegraphics[width=\linewidth]{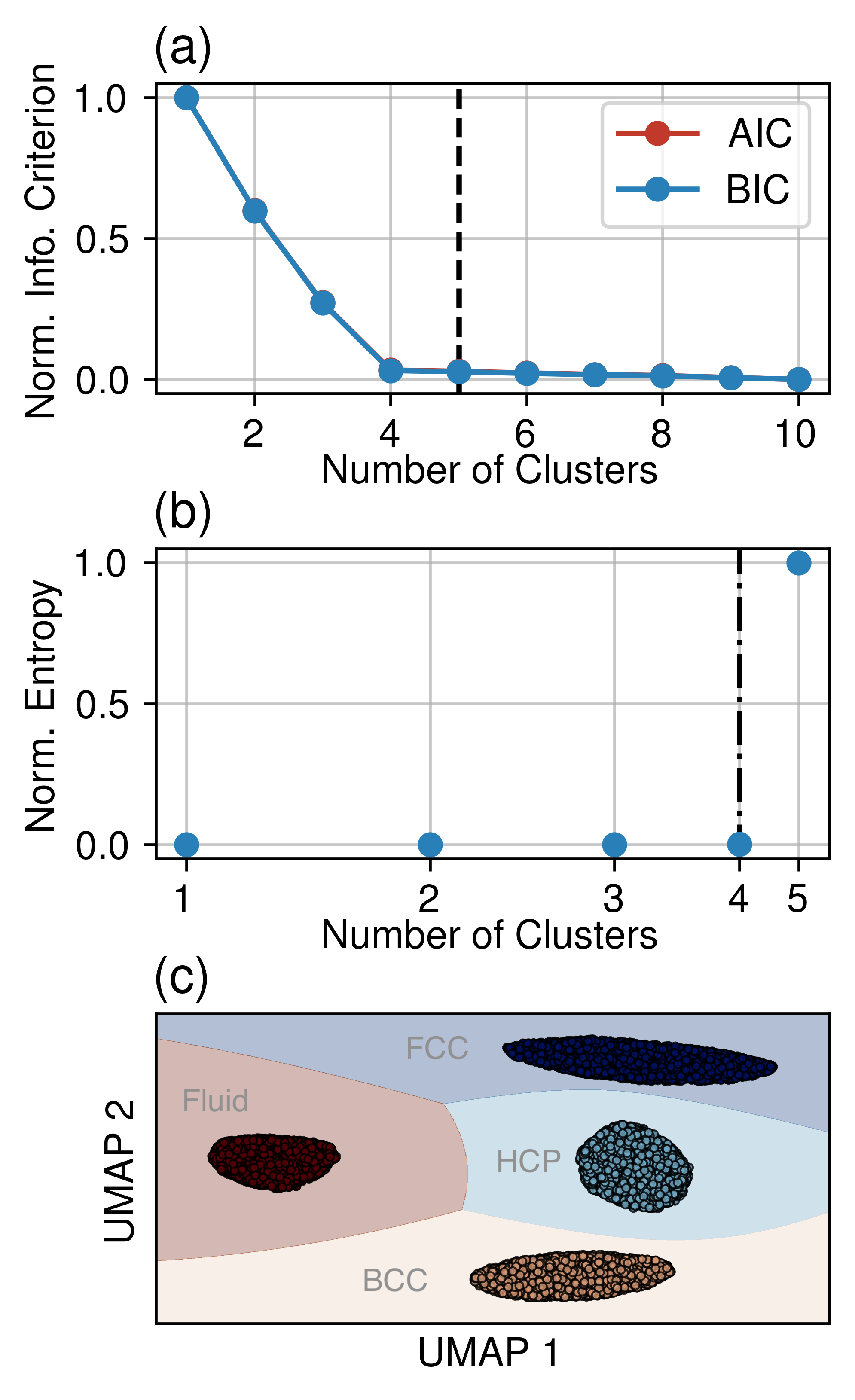}
\caption{Clustering of bulk structures in the UMAP plane. 
    (a) AIC and BIC scores for GMM clustering. (b) Entropy as a function of the number of clusters after successive merging.
    (c) Projection and classification after entropy-based merging into $4$ clusters.}
\label{fig:bulk_umap_criterion}
\end{figure}

Finally, we evaluate the performance of UMAP as a nonlinear dimensionality reduction method for the bulk structural classification. Based on the PCA analysis, we project the data onto a two-dimensional space. After training UMAP using $25$ nearest-neighbors and expected minimum distance $10^{-4}$ as hyperparameters, we apply GMM-based clustering to the resulting low-dimensional embedding, varying the number of components up to 10. As shown in Fig.~\ref{fig:bulk_umap_criterion}(a), both the AIC and BIC scores exhibit a  point of diminishing returns around 4-5 clusters. Here we retain 5 Gaussians.
Following entropy-based merging, the elbow point in the entropy curve occurs at 4 clusters (Fig.~\ref{fig:bulk_umap_criterion}(b)) as expected.

The final classification is shown in Fig.~\ref{fig:bulk_umap_criterion}(c). Notably, UMAP achieves an essentially perfect separation of all four phases: each cluster is extremely compact, well-separated from the others, and free from visible overlap. The fluid, FCC, HCP, and BCC phases each form distinct islands in the UMAP embedding, with clear boundaries and minimal mixing.

\subsubsection{Summary of Bulk Structure Classification}

Since we tested each method on labeled data here (data arising from simulations of the individual phases), we have access to model accuracies. 

\begin{figure}
\centering
\includegraphics[width=\linewidth]{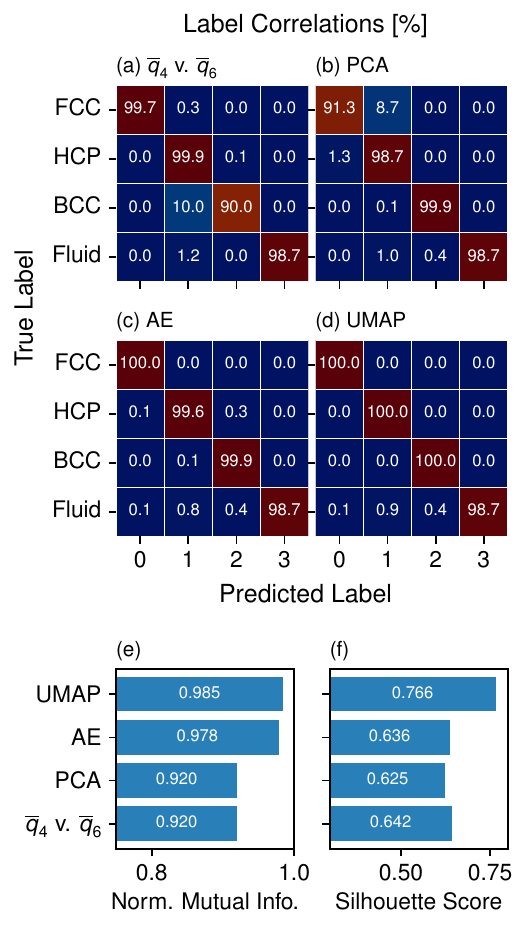}
\caption{(a–d) Confusion matrices showing the correlation between true and predicted labels for each method.(e) Normalized mutual information (NMI) scores and (f) silhouette scores for each dimensionality reduction method for bulk structures . }
\label{fig:bulk_classification_results}
\end{figure}

As a first way to estimate the accuracy, we calculate the confusion matrices, which list for each true label (FCC, HCP, BCC, or fluid) the fraction of samples predicted to be in cluster 0, 1, 2, or 3. For convenience, we have ordered the clusters to be in the same order as the known phases (FCC, HCP, BCC, fluid). The confusion matrices are shown in Fig.~\ref{fig:bulk_classification_results}(a–d), where each row sums to $100\%$. The UMAP-based classification (Fig.~\ref{fig:bulk_classification_results}(d)) is nearly perfect: each crystalline structure forms a pure cluster, with only minor mixing observed in the fluid phase. The autoencoder (Fig.~\ref{fig:bulk_classification_results}(c)) also performs very well, achieving a pure FCC cluster but exhibiting slight mixing for HCP, BCC, and fluid particles.
In contrast, the PCA classification (Fig.~\ref{fig:bulk_classification_results}(b)) shows noticeable mixing between FCC and HCP phases, consistent with the overlap observed in the PCA projections. Similarly, the $\overline{q}_4$ vs. $\overline{q}_6$ classification (Fig.~\ref{fig:bulk_classification_results}(a)) exhibits significant mixing between the BCC and HCP phases, again consistent with Fig. \ref{fig:q4q6aicbic}(c).

As an additional measure to these confusion matrices, we calculate the normalized mutual information (NMI) between the predicted cluster labels $\mathcal{C}$ and the true structural labels $\mathcal{L}$. 
The NMI measures the agreement between the predicted clusters and the true structural labels, normalized to lie between $0$ (no mutual information) and $1$ (perfect agreement). The NMI summarizes the full structure of the confusion matrix and is invariant to permutations of cluster labels. The NMI is defined as:
\begin{equation}
    \mathrm{NMI}(\mathcal{C},\mathcal{L}) = 2 \frac{H(\mathcal{C})-H(\mathcal{C}|\mathcal{L})}{H(\mathcal{C}) + H(\mathcal{L})},
\end{equation}
where $H(\mathcal{C})$ is the information entropy of the predicted clusters and $H(\mathcal{C}|\mathcal{L})$ is the conditional information entropy of the clusters given the true labels:
\begin{eqnarray}
    H(\mathcal{C}) &=& -\sum_{I=1}^C p_\mathcal{C}(I) \log_2(p_\mathcal{C}(I)) \\
    H(\mathcal{C}|\mathcal{L}) &=& -\sum_{I=1}^C\sum_{J=1}^L p_{\mathcal{C}|\mathcal{L}}(I,J) \log_2(p_{\mathcal{C}|\mathcal{L}}(I,J)) 
\end{eqnarray}
The probabilities are given by:
\begin{equation*}
    p_\mathcal{C}(I) = \frac{|C_I|}{N},\quad 
    p_{\mathcal{C}|\mathcal{L}}(I,J) = \frac{|C_I\cap L_J|}{|L_J|}
\end{equation*}
where $C_I$ is the set of particles in cluster $I$, $L_J$ is the set of particles in true label class \( J \), and $N$ is the total number of particles.

As shown in Fig.~\ref{fig:bulk_classification_results}(e), UMAP achieves the highest NMI score ($0.985$), closely followed by autoencoders ($0.978$). PCA and the conventional $\overline{q}_4$ vs. $\overline{q}_6$ method both obtain lower and comparable NMI scores ($0.920$), sharing third place. This ranking is consistent with our qualitative observations from the cluster plots.

Finally, to quantify to what extent the dimensionality reduction results in well-separated clusters, we calculate the silhouette score \cite{rousseeuw1987silhouettes}. For a given particle $i$, the silhouette score is defined as the normalized difference between the mean intra-cluster distance $a(i)$ and the mean nearest-cluster distance $b(i)$:
\begin{equation}
    s(i) = \frac{b(i)-a(i)}{\max\{a(i),b(i)\}}
\end{equation}
where
\begin{equation*}
    a(i) = \frac{1}{|C_I|-1} \sum_{j \in C_I} d(i,j), \quad
    b(i) = \min_{J \neq I} \frac{1}{|C_J|} \sum_{j \in C_J} d(i,j),
\end{equation*}
and $C_I$ is the cluster to which particle $i$ belongs, and $d(i,j)$ is the distance between particles $i$ and $j$ in the dimension-reduced space.

A silhouette score close to $1$ indicates that the particle is well-matched to its own cluster and poorly matched to neighboring clusters, while values near 0 suggest ambiguity in assignment, and negative values indicate potential misclassification. For the entire system, the silhouette score is then defined as an average over all particles. Note that unlike the confusion matrix and NMI, the silhouette score can be calculated even for unlabeled data.

The dataset averaged silhouette scores are presented in Fig.~\ref{fig:bulk_classification_results}(f). Here, UMAP clearly outperforms the other three methods, as it exhibits the highest degree of cluster separation ($0.766$), while the other three methods achieve approximately equal scores $\sim 0.63$.

In summary, UMAP clearly outperforms the other methods in both classification accuracy and cluster separability, with autoencoders following closely behind. PCA and the conventional $\overline{q}_4$ vs. $\overline{q}_6$ analysis, while capable of distinguishing major features, struggle to cleanly separate closely related crystal structures.

\subsection{Icosahedral Supraparticle}

\begin{figure}
    \centering
    \includegraphics[width=0.7\linewidth]{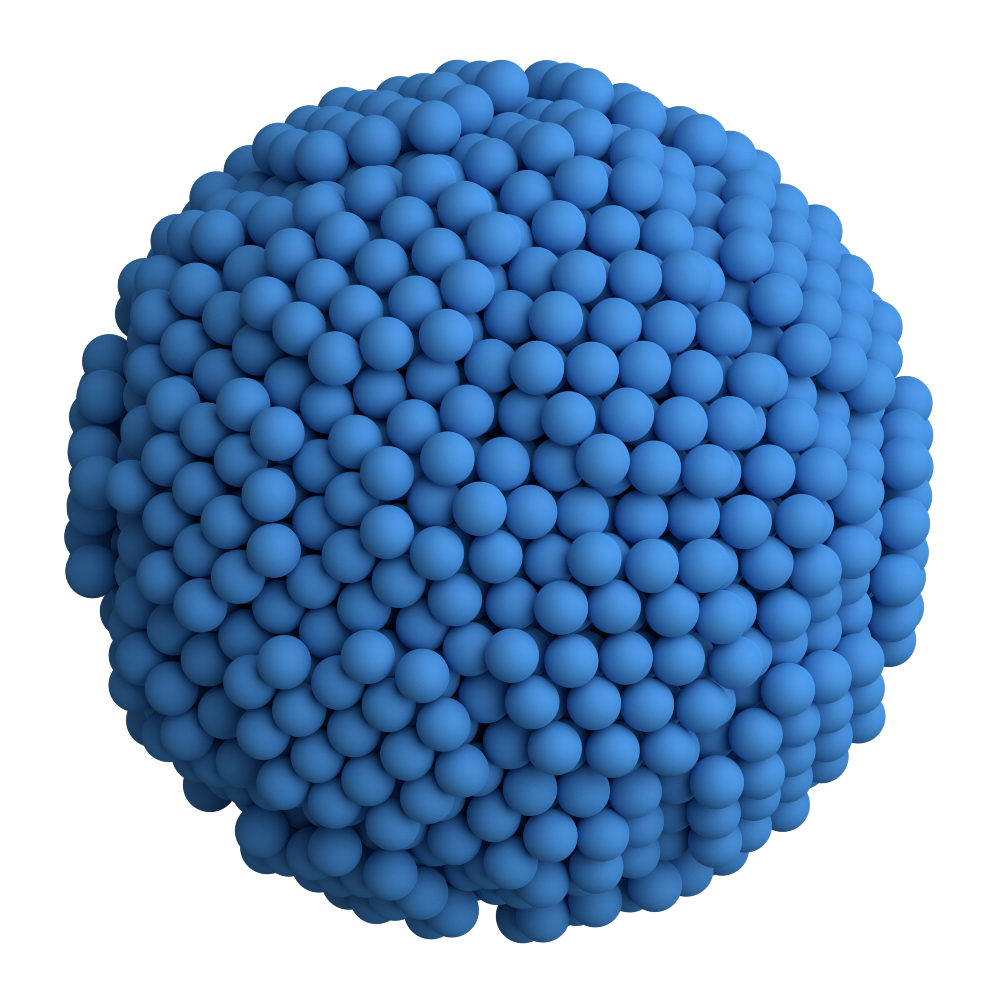}
    \caption{One of the configurations used in the self-assembled icosahedral supraparticle dataset. Radially cut to highlight the structure.}
    \label{fig:supraparticle}
\end{figure}

We now test the same methods in a much more complex scenario, where many different local environments are possible. Specifically, we look at configurations of hard spheres in spherical confinement, which are known to form complex icosahedral clusters (see e.g. Refs. \onlinecite{de2015entropy,buckmann2025quantitative}) that contain e.g. crystal-like domains as well as planar and linear features that could be significantly harder to detect due to their smaller size. An example of such a cluster is shown in Fig. \ref{fig:supraparticle}. 

To this end, we prepare a second dataset by performing 12 event-driven molecular dynamics (EDMD) simulations of hard spheres~\cite{smallenburg2022efficient} under spherical confinement using hard-wall boundary conditions. The number of particles is varied between $100$ and $8000$, with the precise particle counts chosen according to magic numbers~\cite{mackay1962dense,wang2018magic} known to favor the formation of structures with icosahedral symmetry. Each simulation is initialized at a packing fraction of $0.3$, and the spherical confinement is gradually shrunk at a constant speed until a jammed configuration is reached. For each simulation, we extract two snapshots: one at packing fraction $0.57$ and another at $0.56$. Subsequently, we calculate the local structure descriptors for each particle as described in Subsection~\ref{subsec:descriptors}, resulting in a combined dataset comprising $78\ 184$ particles in total.

Following the procedure established for the bulk crystalline structures, we apply dimensionality reduction and clustering to this dataset. In contrast to the bulk systems, where the true phase labels were known, no \emph{a priori} labeling is available here. As such, the resulting clusters must be interpreted based solely on the emergent organization in the low-dimensional embeddings, and on visual inspection of the resulting configuration.

\begin{figure}
\centering
\includegraphics[width=\linewidth]{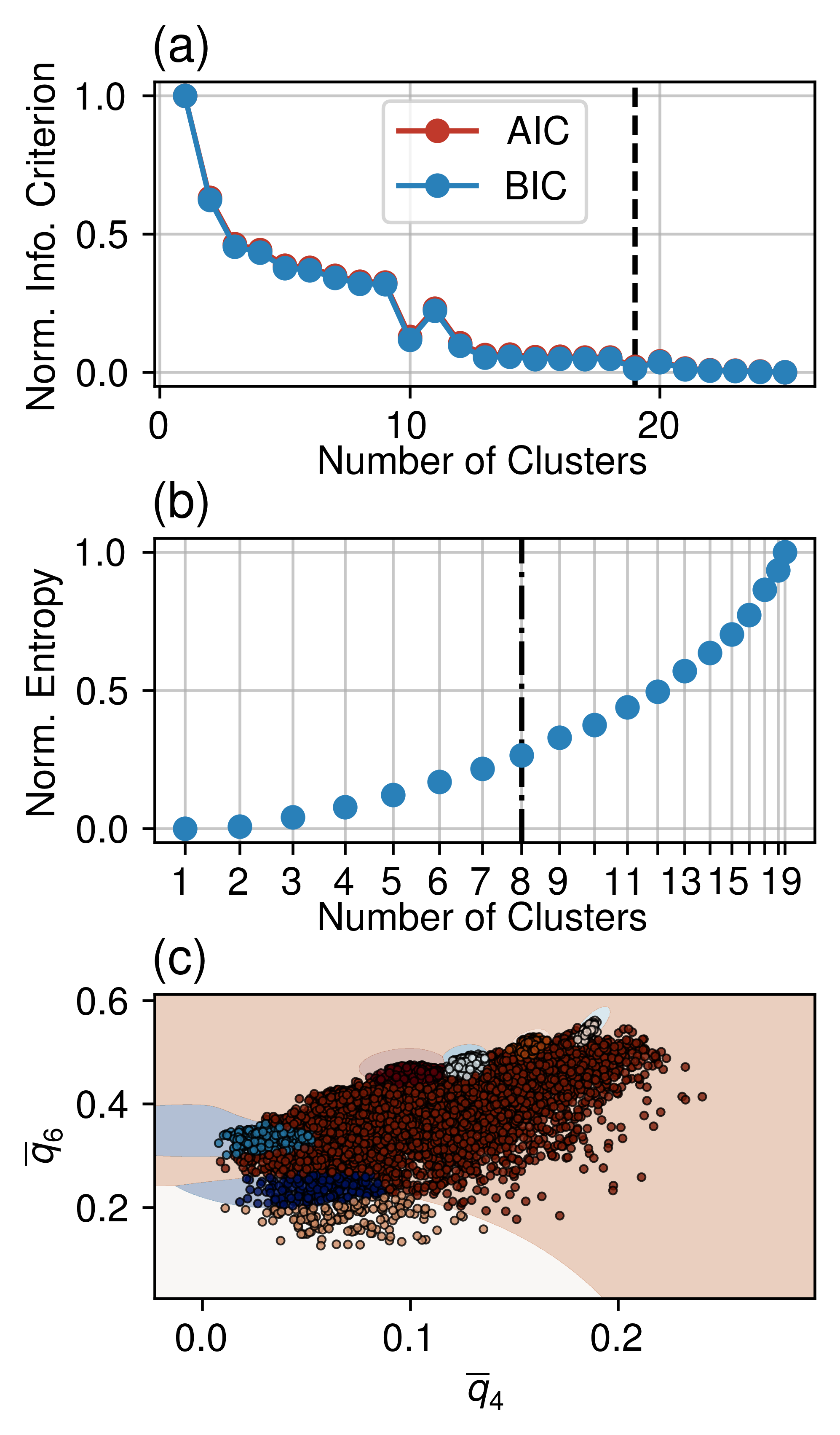}
\caption{Clustering of icosahedral supraparticles in the $\overline{q}_4$ vs. $\overline{q}_6$ plane. 
    (a) AIC and BIC scores for GMM clustering. (b) Entropy as a function of the number of clusters after successive merging.
    (c) Projection and classification after entropy-based merging into $8$ clusters.}
\label{fig:mnc_q4q6_criterions}
\end{figure}

As with the bulk crystalline dataset, we first establish a baseline classification by clustering the supraparticle dataset in the $\overline{q}_4$ vs. $\overline{q}_6$ plane. Gaussian mixture models are trained with up to $25$ components, and model selection is performed using the AIC and BIC criteria. As shown in Fig.~\ref{fig:mnc_q4q6_criterions}(a), both AIC and BIC exhibit a shallow minimum at approximately 19 components. This suggests a relatively complex underlying structure, reflecting the heterogeneous local environments expected in confined self-assembly.

Unlike in the bulk crystalline case, the entropy curve, shown in Fig.~\ref{fig:mnc_q4q6_criterions}(b), does not display a sharp elbow. Instead, it decreases gradually as the number of clusters decreases, making the identification of a natural number of clusters more ambiguous. We (somewhat arbitrarily)  select $8$ clusters for the final classification, balancing model simplicity with the underlying structural richness of the dataset.

The final clustering is visualized in Fig.~\ref{fig:mnc_q4q6_criterions}(c), where the particles are colored according to their assigned cluster in the $\overline{q}_4$ vs. $\overline{q}_6$ plane. Clearly, this characterization does not result in a set of clearly separated clusters.

\subsubsection{Principal component analysis}

\begin{figure}
\centering
\includegraphics[width=\linewidth]{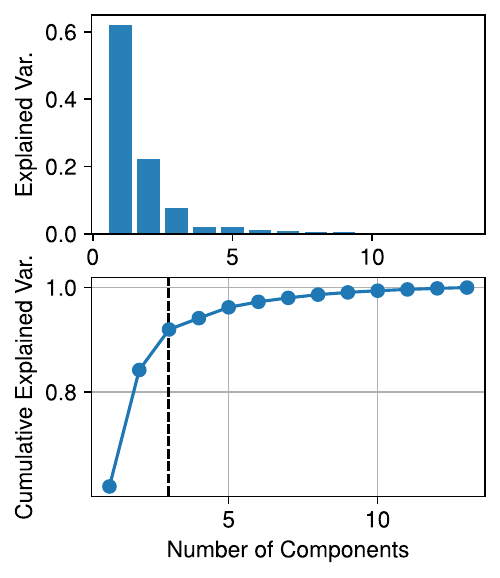}
\caption{Dimensionality analysis of spherically confined hard spheres. (a) Variance explained by individual principal components. (b) Cumulative explained variance.}
\label{fig:mnc_pca_dim}
\end{figure}

We next apply PCA to the supraparticle dataset.
The variance explained by each principal component, shown in Fig.\ref{fig:mnc_pca_dim}(a), indicates that the first three principal components capture a substantial fraction of the total variance. The cumulative explained variance curve in Fig.\ref{fig:mnc_pca_dim}(b) further confirms that projecting onto three dimensions retains over $90\%$ of the structural information. Therefore, we project the data into a three-dimensional principal component space for subsequent analysis.

\begin{figure}
\centering
\includegraphics[width=\linewidth]{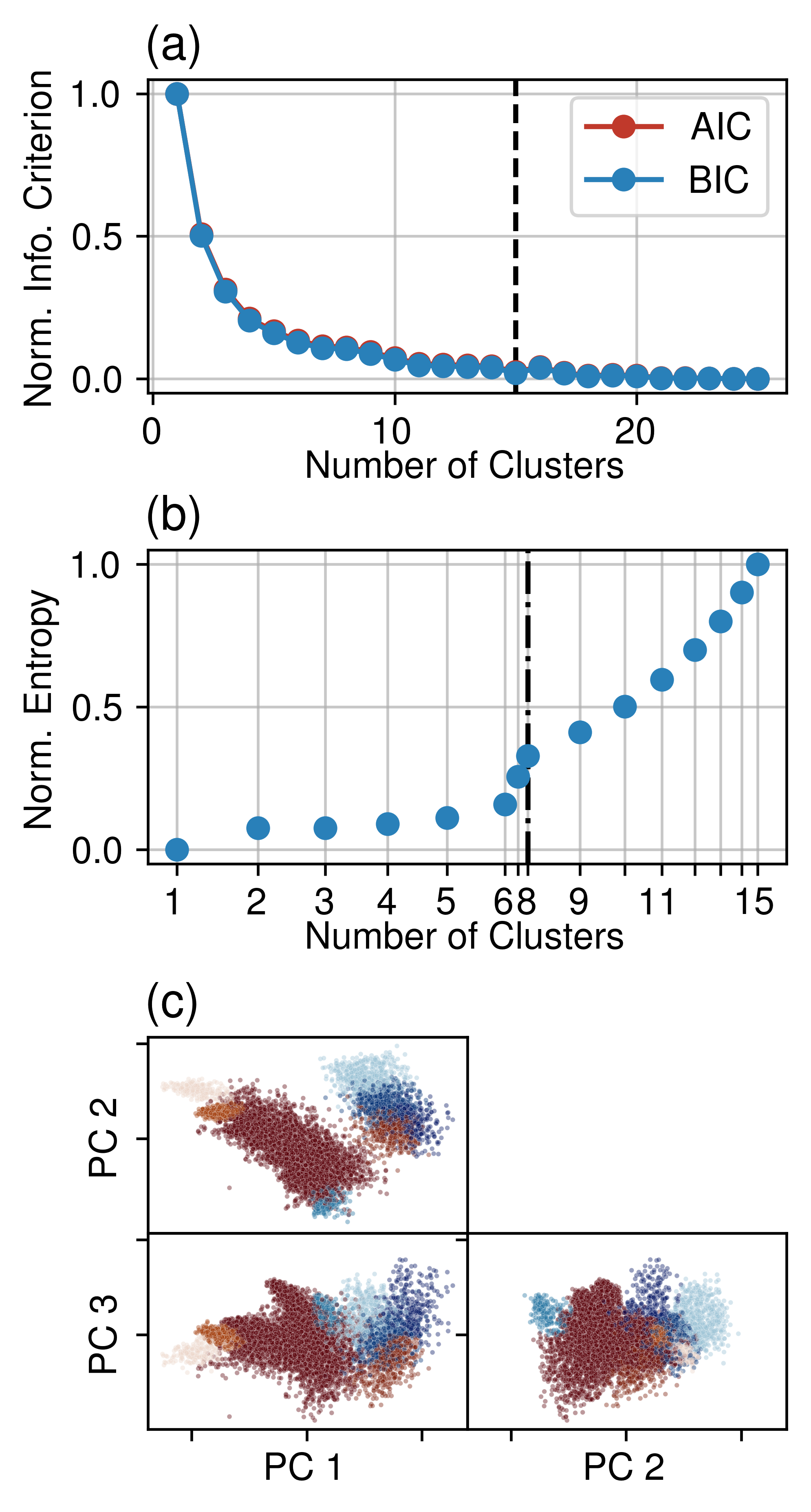}
\caption{Clustering of icosahedral supraparticles in the PCA projection. 
    (a) AIC and BIC scores for GMM clustering. (b) Entropy as a function of the number of clusters after successive merging.
    (c) Projection and classification after entropy-based merging into $8$ clusters.}
\label{fig:mnc_pca_criterions}
\end{figure}

We then apply GMM-based clustering to the PCA-reduced data, training models with up to 20 components. As shown in Fig.~\ref{fig:mnc_pca_criterions}(a), both the AIC and BIC decay as expected.  We choose 15 Gaussians for our initial model. 
As in the previous cases, we perform entropy-based merging to obtain a more interpretable clustering. The entropy curve, shown in Fig.~\ref{fig:mnc_pca_criterions}(b), exhibits a sharp transition rather than an elbow. This transition occurs between $6$--$8$ clusters. After visual inspection of the clustering, we select $8$ as the final number of clusters to capture the major structural diversity present.

The final clustering results are shown in Fig.~\ref{fig:mnc_pca_criterions}(c), visualized by plotting the projections onto the first three principal components. Multiple distinct groups are evident, although significant overlap remains between several clusters.

\subsubsection{Autoencoders}

We next apply a neural network-based autoencoder to perform nonlinear dimensionality reduction on the supraparticle dataset. Following the PCA analysis, we design the autoencoder to project the structural descriptors onto a three-dimensional latent space.

\begin{figure}
\centering
\includegraphics[width=\linewidth]{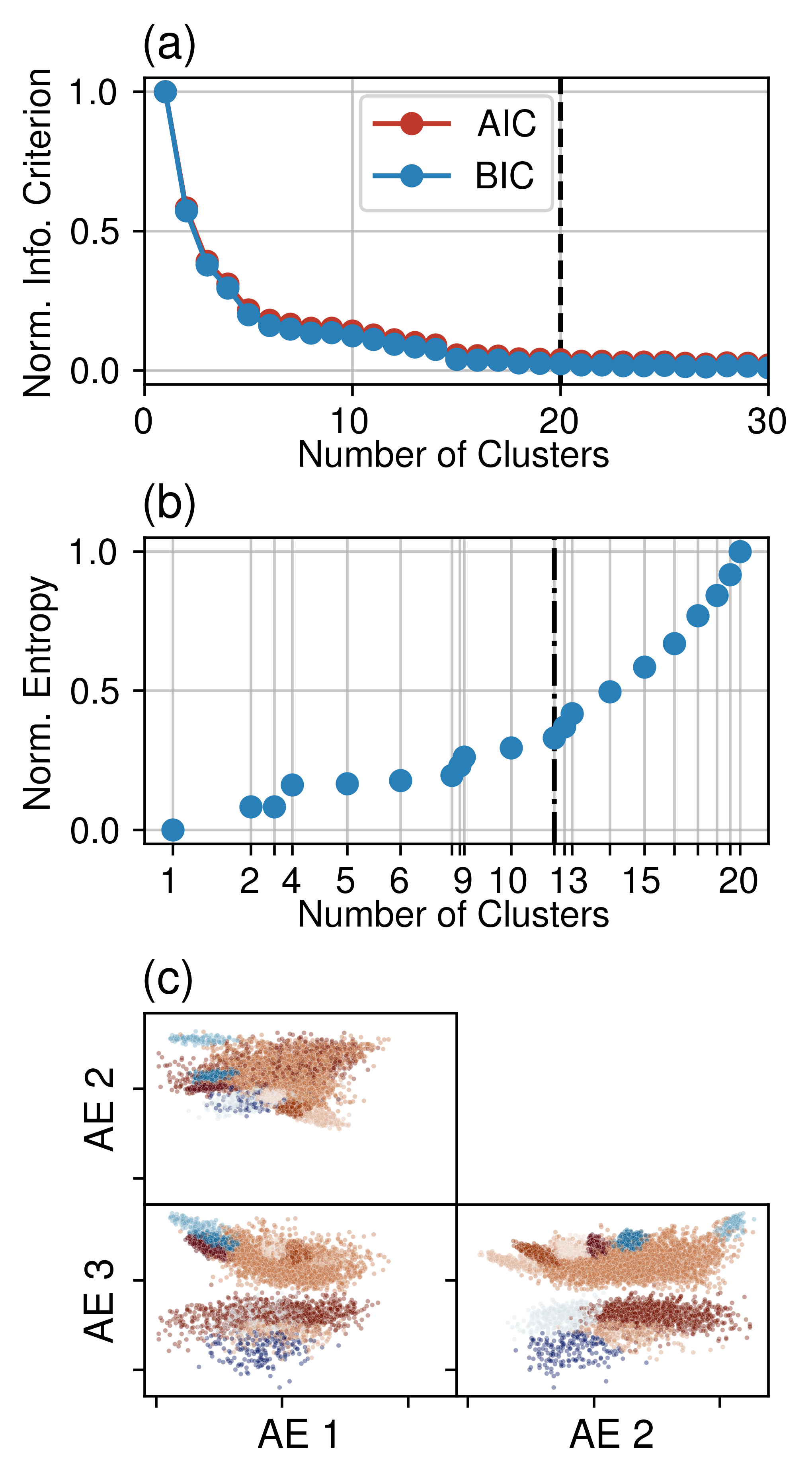}
\caption{Clustering of icosahedral supraparticles in the autoencoder projection. 
    (a) AIC and BIC scores for GMM clustering. (b) Entropy as a function of the number of clusters after successive merging.
    (c) Projection and classification after entropy-based merging into $11$ clusters.}
\label{fig:mnc_ae_criterions}
\end{figure}

\begin{figure}
\centering
\includegraphics[width=\linewidth]{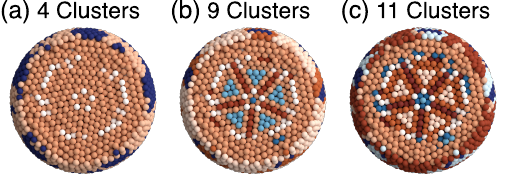}
\caption{Entropy-based cluster merging of an icosahedral supraparticle using the autoencoder projection into (a) $4$, (b) $9$, (c) $11$ clusters.}
\label{fig:mnc_ae_series}
\end{figure}

After training the autoencoder, we project the data into the latent space and apply GMM-based clustering. As shown in Fig.~\ref{fig:mnc_ae_criterions}(a), both the AIC and BIC scores flatten after $15$ clusters. To give more flexibility to cluster merging, we choose $20$ as the initial number of clusters.
We then apply entropy-based merging to simplify the clustering. The entropy curve, shown in Fig.~\ref{fig:mnc_ae_criterions}(b), exhibits multiple noticeable transitions rather than a single sharp elbow. Specifically, we observe three distinct changes in slope occurring at approximately $4$, $9$, and $11$ clusters.
 Upon visual inspection of the corresponding cluster structures in Fig.~\ref{fig:mnc_ae_series}, these transition points align with meaningful coarse-graining steps in the organization of the supraparticle. 
 At 4 clusters, the system primarily distinguishes surface particles from bulk particles. 
 At 9 clusters, the first shell of particles beyond the surface becomes resolved as a distinct structural domain. 
 Finally, at 11 clusters, finer substructures within the core of the supraparticle are further differentiated.

To preserve as much structural detail as possible without overfitting, we select the transition point corresponding to $11$ clusters. 
The final classification is visualized in Fig.~\ref{fig:mnc_ae_criterions}(c), plotted in the three-dimensional autoencoder latent space. While some overlap persists between clusters, they appear to overlap significantly less than we observed for PCA. 

\subsubsection{Uniform Manifold Approximation and Projection}

We now apply UMAP to the supraparticle dataset using the same hyperparameters that we use for the bulk structures, and again project onto three dimensions. 

\begin{figure}
\centering
\includegraphics[width=\linewidth]{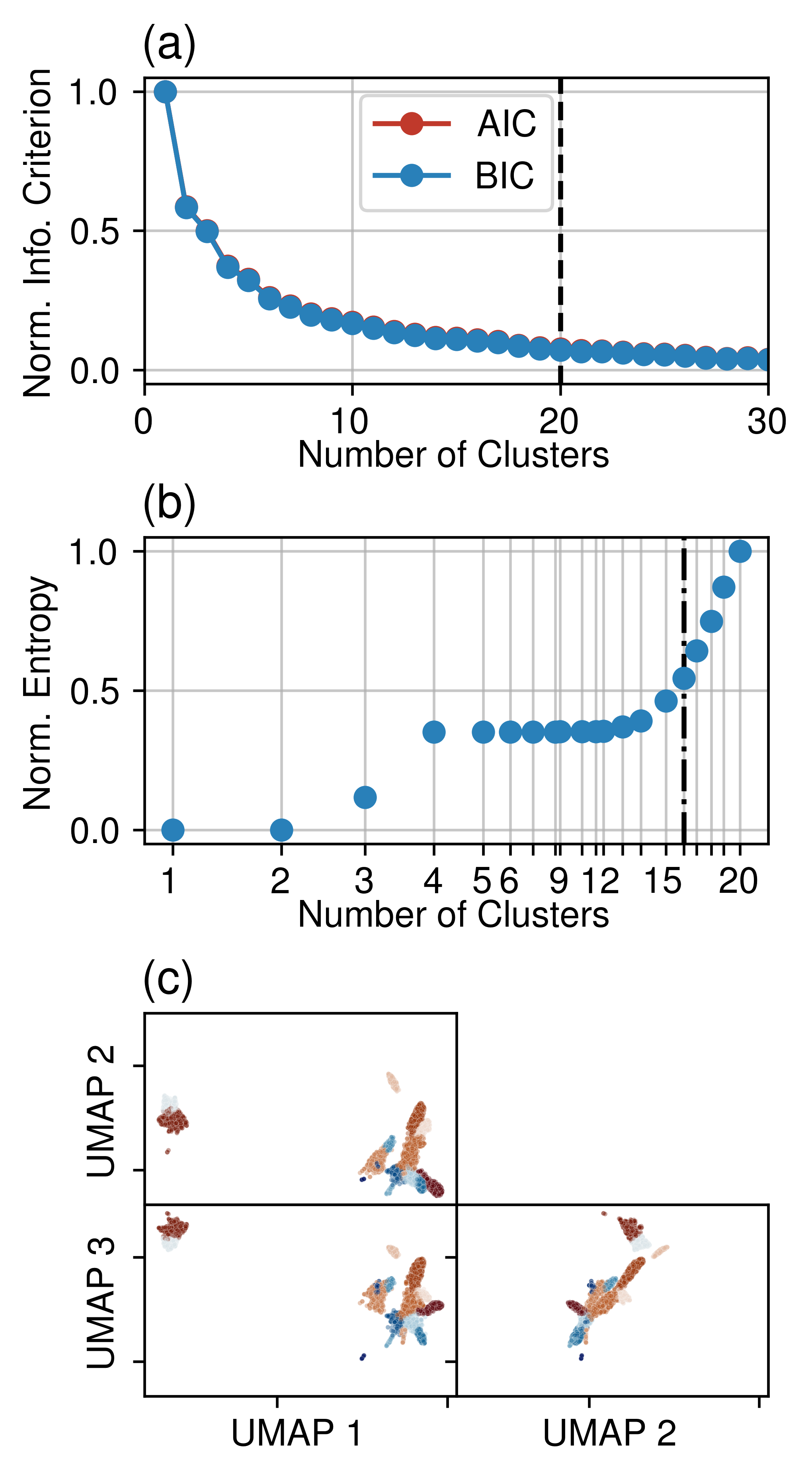}
\caption{Clustering of icosahedral supraparticles in the UMAP. 
    (a) AIC and BIC scores for GMM clustering. (b) Entropy as a function of the number of clusters after successive merging.
    (c) Projection and classification after entropy-based merging into $16$ clusters.}
\label{fig:mnc_umap_criterions}
\end{figure}

The GMM clustering procedure is shown in Fig.~\ref{fig:mnc_umap_criterions}(a), where both the AIC and BIC scores decay continuously. Starting from 20 clusters in the cluster merging procedure, the entropy curve, shown in Fig.~\ref{fig:mnc_umap_criterions}(b), reveals two major transition points: the first at 4 clusters and the second at $12$--$16$ clusters.

\begin{figure}
\centering
\includegraphics[width=\linewidth]{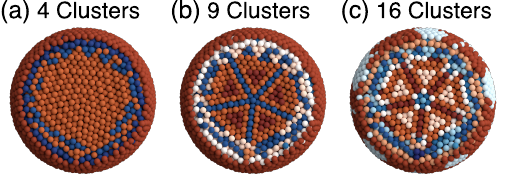}
\caption{Entropy-based cluster merging of an icosahedral supraparticle using the UMAP into (a) $4$, (b) $9$, (c) $16$ clusters.}
\label{fig:mnc_umap_series}
\end{figure}

Visual inspection of the cluster structures associated with these points reveals clear physical interpretations as seen in Fig.~\ref{fig:mnc_umap_series}. At 4 clusters, the clustering distinguishes broadly between surface particles, first-layer particles, and core particles. While, between 4 and 16 clusters, the clustering progressively resolves substructures within these macroscale regions.

We therefore select $16$ clusters as the final number for classification, balancing resolution of important substructures with avoidance of overfitting.
The final clustering is visualized in Fig.~\ref{fig:mnc_umap_criterions}(c), plotted in the three-dimensional UMAP-reduced space. Compared to PCA and autoencoders, the UMAP embedding achieves significantly better cluster separation. While some overlap persists, the overall clarity of the cluster organization is markedly improved.

\subsubsection{Summary of Supraparticle Structure Classification}

To quantitatively evaluate the performance of the different dimensionality reduction methods on the supraparticle dataset, we compute the silhouette score for each method. Recall that the silhouette score quantifies the degree of separation between clusters, with higher values indicating better cluster compactness and separation.

\begin{figure}
\centering
\includegraphics[width=\linewidth]{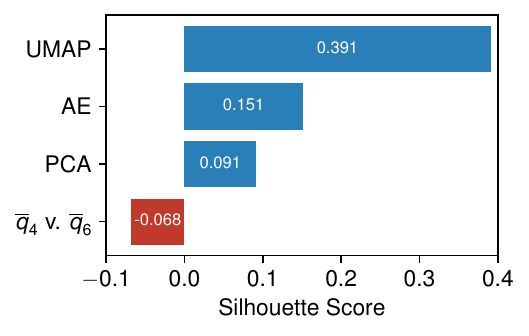}
\caption{Silhouette scores for clustering in the supraparticle dataset for each method. Higher values indicate better cluster separation.}
\label{fig:mnc_silhouette}
\end{figure}

As shown in Fig.~\ref{fig:mnc_silhouette}, none of the silhouette scores are extremely high. However, UMAP achieves the highest score (0.391), followed by the autoencoder ($0.151$), and PCA (0.091). The $\overline{q}_4$ vs. $\overline{q}_6$ approach performs significantly worse, with a negative silhouette score ($-0.068$). These results confirm that UMAP provides the clearest structural separation among the methods tested, while linear techniques struggle to resolve the complexity of the supraparticle assembly.

\begin{figure*}
    \centering
    \includegraphics[width=\textwidth]{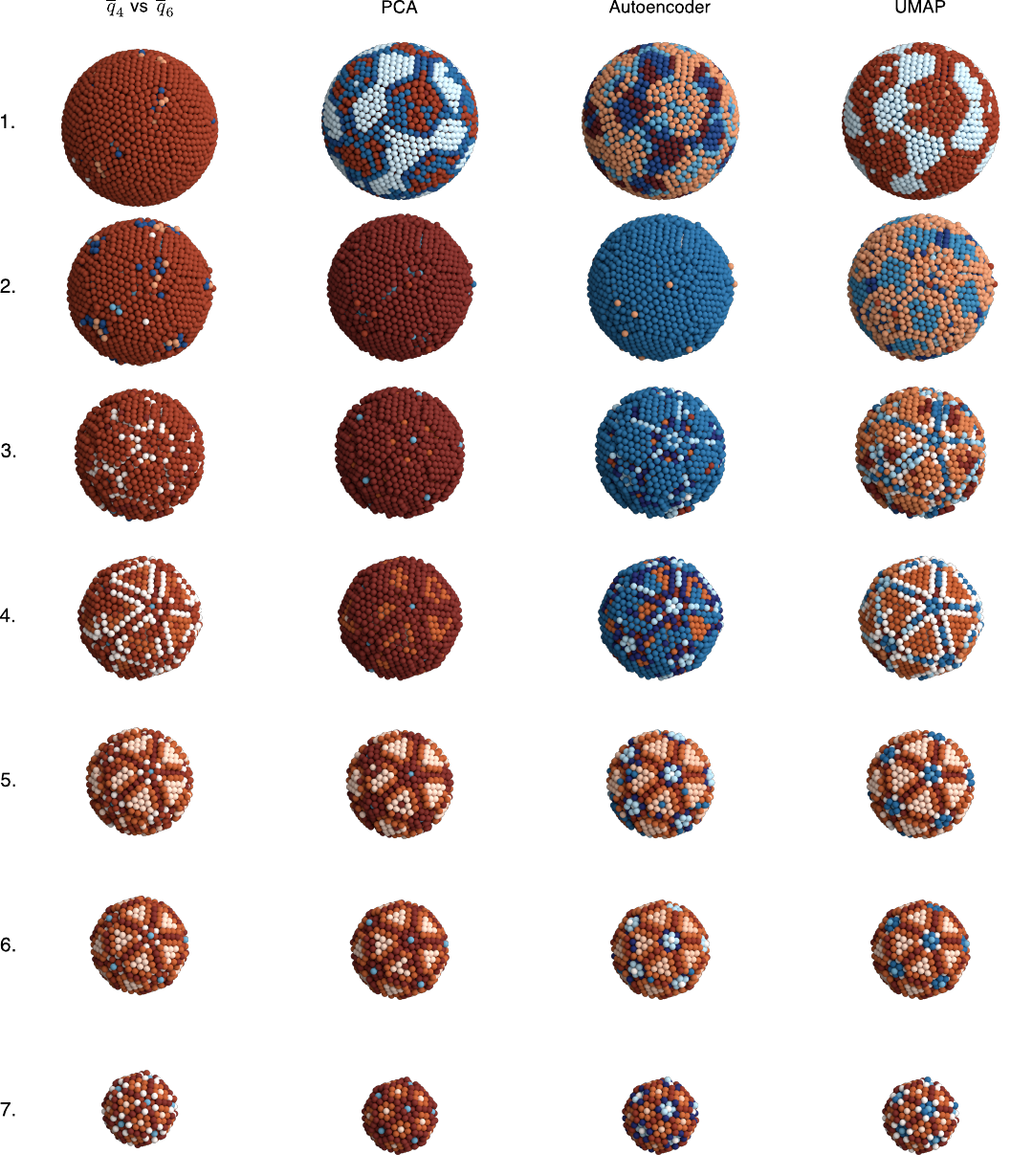}
    \caption{Final cluster assignments for each method (columns) across radial cuts of the same supraparticle configuration (rows).}
    \label{fig:mncclusters}
\end{figure*}
    
\begin{figure}
    \centering
    \includegraphics[width=\linewidth]{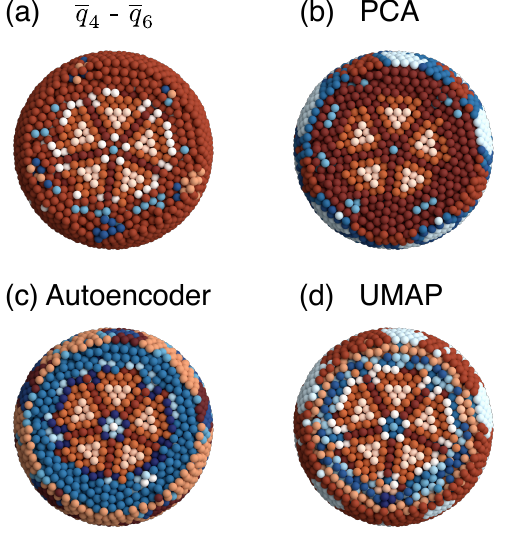}
    \caption{Final cluster assignments for each method shown by a planer cut through a $5$-fold symmetry center of the same supraparticle configuration.}
    \label{fig:planecut}
\end{figure}

For a final comparison between the different approaches, we look at the classified configurations and examine how the classifications differ in different regions of the suprastructure. The results are shown in Figs.~\ref{fig:mncclusters} and \ref{fig:planecut}. In Fig. \ref{fig:mncclusters}, each column represents a different method and each row corresponds to a radial slice through the same $7500$-particle supraparticle configuration. In Fig. \ref{fig:planecut}, we additionally show for each method a planar cut through the same $5$-fold symmetry center of the supraparticle.

For the $\overline{q}_4$ vs. $\overline{q}_6$ method, the classification is poor in the outer regions. In the outer shell, nearly all particles are assigned to the same cluster, indicating that this descriptor set lacks the resolution to distinguish structural variations at the surface. In the second and third row of Fig. \ref{fig:mncclusters}, i.e., a few layers below the surface, this cluster persists across most of the surface, with only faint signs of emerging structure at 5-fold symmetry centers. From the fourth row inward, deeper into the supraparticle, the core begins to emerge with clearer boundaries. Notably, from rows five through seven, the core becomes well-organized into distinct domains: each tetrahedral unit appears divided into a central core and a surrounding shell, and additional clusters capture particles located at the interface between tetrahedra as well as a tube-like structure aligned with the 5-fold symmetry axes.

In the PCA column, we observe improved sensitivity at the surface. The first row of Fig. \ref{fig:mncclusters} reveals many more structural domains, particularly in the 5-fold symmetric regions, which appear as well-defined pentagonal clusters, distinct from the interstitial surface domains. However, these pentagonal regions are split into three subclusters, likely over-fragmenting what should be a single class. The second and third rows largely resemble the behavior of $\overline{q}_4$ vs. $\overline{q}_6$, though the third row more successfully isolates the 5-fold centers. In the fourth row, we begin to transition to the more clearly classified cluster core, which becomes clearly visible in the fifth through seventh rows. Interestingly, in the fifth and sixth rows, PCA appears to perform worse than $\overline{q}_4$ vs. $\overline{q}_6$: the intermediate clusters at tetrahedral interfaces and tubular regions are not clearly resolved, and these particles are instead absorbed into the surrounding shell cluster.

The autoencoder again performs similarly to PCA in the outer few shells. However, at deeper levels (third and following rows of Fig. \ref{fig:mncclusters}), its classification picks out more detail than both PCA and $\overline{q}_4$ vs. $\overline{q}_6$. In particular, we see that in the core, the autoencoder is capable of distinguishing more distinct environments around the five-fold symmetry centers.

Finally, UMAP provides the best performance across all radial layers. In the first row, the surface particles are immediately divided into distinct clusters, separating the 5-fold symmetric regions from the remainder of the surface. In the second row, UMAP uniquely resolves the structure into two structurally distinct clusters, consistent with the structure found at the cluster surface. In the third row, an additional cluster emerges to identify a transition layer i.e., a structural buffer between the surface shells and the densely packed core. From the fourth row onward, the core regions are robustly classified. The organization of tetrahedral units, including their cores and interfacial zones, is consistently resolved. In the fifth and sixth rows, UMAP retains the fine structure seen with the autoencoder, but with crisper boundaries and less cluster mixing, leading to the cleanest classification of the supraparticle’s full radial and angular structure.

In summary, while all methods are capable of revealing significant aspects of the complex internal structure of the supraparticle, UMAP appears to perform the best. It captures not only the subtle variations in structure in the surface layers, as well as the intricate variations inside the more ordered core. 

\subsection{Test of classification on an experimental icosehedral supraparticle  }

\begin{figure}
    \centering
    \includegraphics[width=\linewidth]{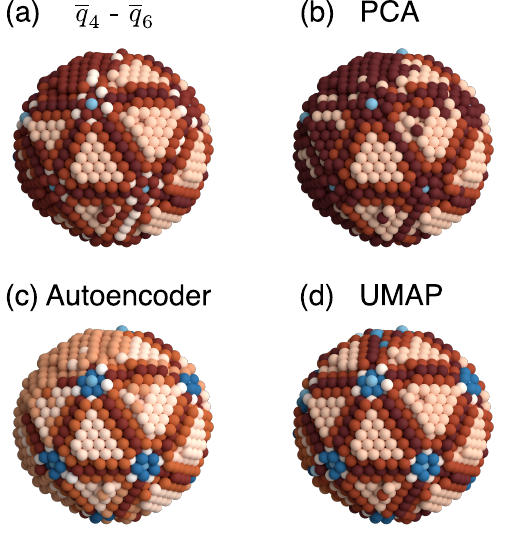}
    \caption{The cluster assignments of an experimental icosahedral supraparticle. The structure is radially cut to reveal the inner structure.}
    \label{fig:expsupra}
\end{figure}

As a last test, we use the UMAP algorithm to study the structure of a supraparticle formed experimentally. In  Fig. \ref{fig:expsupra}, we show the central part of the experimental supraparticle. Similar to the simulation data, the result associated with the UMAP clustering are the most robust.  More specifically, it is the only one of the three that accurately captures the structure on the top left of the supraparticle. Moreover, it generally shows fewer structural misclassifications.

\section{Conclusions}\label{sec:discussion}


In this work, we examined how different dimensionality-reduction approaches influence the performance of a common unsupervised pipeline for identifying structural motifs in crystalline and self-assembled systems. Our descriptor set consists of the averaged bond-orientational order parameters ($\ell =1,\ldots,12$),  together with a scalar that quantifies the deviation of a particle’s position from the center of mass of its neighbors. As a baseline embedding, we used 4- vs. 6-fold symmetric averaged bond orientational order parameters ($\overline{q}_4$ vs. $\overline{q}_6$) directly. We then compared it to three alternative dimensionality-reduction methods: principal component analysis (PCA), autoencoders (AE), and uniform manifold approximation and projection (UMAP). All embeddings were analyzed using the same clustering model, namely Gaussian mixture models with entropy-based merging, which allows us to isolate the influence of the embedding method itself. We apply this workflow to both labeled bulk crystalline phases and unlabeled icosahedral supraparticle assemblies.

For the bulk crystalline dataset, where ground truth labels are available, we found that UMAP and autoencoders provided the highest classification accuracy, with UMAP achieving the best normalized mutual information and silhouette scores. Both $\overline{q}_4$ vs. $\overline{q}_6$ and PCA recovered coarse structural distinctions but struggled to fully separate similar phases. 

In the more challenging case of icosahedral supraparticles, where no \emph{a priori} labeling is available, UMAP again outperformed the other methods in terms of its ability to robustly distinguish between different local structural features. In particular, UMAP was more capable than the other methods at recognizing subtle structural variations in the outer layers of the supraparticle, where significant disorder is present. Additionally, inside the more ordered core, it was able to robustly identify more distinct structural variations. 

An important consideration in the deployment of these methods is the complexity of hyperparameter tuning. PCA and $\overline{q}_4$ vs. $\overline{q}_6$ analysis require no tuning, offering simplicity at the expense of expressiveness. UMAP, by contrast, has only two hyperparameters: the number of neighbors and the minimum distance between embedded points. Of these, the minimum distance primarily affects the visual compactness of the embedding without significantly altering the graph structure, making UMAP effectively a single-parameter model. Despite this minimal complexity, UMAP consistently yields superior results. Autoencoders, while also capable of high performance, require careful tuning across a large hyperparameter space making them more cumbersome to deploy and more susceptible to overfitting.

In conclusion, unsupervised learning, when paired with physically informed descriptors, offers a robust framework for structure classification in both simple and complex systems. Among the dimensionality-reduction methods tested, UMAP provides the best trade-off between accuracy, interpretability, and algorithmic simplicity. This makes UMAP particularly well-suited for unsupervised discovery of structural classes in condensed matter systems.

\section{Supplementary Material}
In the supplementary material, we systematically explore the effect of additional hidden layers on the autoencoder approach.

\section{Acknowledgments}
We thank Marjolein de Jager for providing the bulk crystalline snapshots and Stefanie D. Pritzl for fruitful discussions. AU, RY, AvB and LF acknowledge funding from the
Dutch Research Council (NWO) under the grant number OCENW.GROOT.2019.071. JIB did this work as part of the Advanced Research Center Chemical Building Blocks Consortium, ARC CBBC, which is cofounded and cofinanced by the Dutch Research Council (NWO) and The Netherlands Ministry of Economic Affairs and Climate Policy.
LDH acknowledges funding from the Netherlands Center for Multiscale Catalytic Energy Conversion (MCEC), an NWO Gravitation programme funded by the Ministry of Education, Culture and Science of the government of the Netherlands.

\section{Conflict of Interest}
The authors have no conflicts to disclose.
\section{Data Availability}
Data and code will be published on Zenodo and Github prior to publication of the paper.

\bibliography{refs}

\end{document}